\documentclass[pra,preprintnumbers,tightenlines,twocolumn,superscriptaddress]{revtex4-1}

\usepackage{amsmath}
\usepackage{amssymb}
\usepackage{graphicx}
\usepackage{color}
\usepackage[utf8]{inputenc}

\newcommand{\bra}[1]{\langle #1 |}
\newcommand{\ket}[1]{| #1 \rangle}
\newcommand{\mTuple}[1] {{\mathbf #1}}
  
\newcommand{\figd}[1]{{#1}}
\graphicspath{{{./}}}

\begin{document}

\title{Coherently delocalized states in dipole interacting Rydberg ensembles: the role of internal degeneracies}

\author{Ghassan Abumwis}
\affiliation{Max-Planck-Institut f\"ur Physik komplexer Systeme, N\"othnitzer Str.\ 38, 
D-01187 Dresden, Germany }

\author{Christopher W. W\"achtler}
\affiliation{Max-Planck-Institut f\"ur Physik komplexer Systeme, N\"othnitzer Str.\ 38, 
D-01187 Dresden, Germany }
	
\author{Matthew T. Eiles}
\affiliation{Max-Planck-Institut f\"ur Physik komplexer Systeme, N\"othnitzer Str.\ 38, 
D-01187 Dresden, Germany }

\author{Alexander Eisfeld}
\email{eisfeld@pks.mpg.de}
\affiliation{Max-Planck-Institut f\"ur Physik komplexer Systeme, N\"othnitzer Str.\ 38,
D-01187 Dresden, Germany }

\begin{abstract}
We investigate the effect of degenerate atomic states on the exciton delocalization of dipole-dipole interacting Rydberg assemblies.
Using a frozen gas and regular one-, two-, and three-dimensional lattice arrangements as examples, we see that degeneracies can enhance the delocalization compared to the situation when there is no degeneracy. 
Using the Zeeman splitting provided by a magnetic field, we controllably lift the degeneracy to study in detail the transition between degenerate and non-degenerate regimes. 
\end{abstract}
\maketitle

\section{Introduction}
The formation of states where an electronic excitation is coherently delocalized over several particles plays an important role in many systems. 
Examples include light harvesting in photosynthesis \cite{AmVaGr00__}, molecular aggregates \cite{Ko96__,SaEiVa13_21_}, quantum dot arrays \cite{CrHoTr02_186802_,GeStDo05_137403_}, metallic nanoparticles \cite{QuLeKr98_1331_}, and Rydberg atoms \cite{AnVeGa98_249_,Ga05__,RoHeTo04_42703_,LiTaJa06_27_}.
These coherent collective states are formed by the interaction of transition dipoles of the individual particles.
The relevant transition dipoles connect two eigenstates of a particle with {\it different energy}, and are obtained by evaluating the dipole-operator between the respective states. 
Typically, delocalized states strongly modify the absorption properties and allow an initially localized excitation to be transfered along the assembly of particles.
 
Because of their relevance for light harvesting, the delocalization properties of assemblies of molecules have been extensively studied for many years.
For most molecules of interest, within each constituent atom only two electronic states participate (typically the first electronically excited state and the electronic ground state), resulting in a single relevant transition energy for the molecule. 
However, in all of the systems mentioned above,  the single particle transition energies can become degenerate.
In the molecular case such degenerate transition energies can be due to an underlying symmetry \cite{Go15_259_,VlEi14_305301_,ZhLuZh16_623_,LuChZh19_233901_}; similar symmetry-induced degeneracies are present in quantum dots and metal nano-particles.
The spherical symmetry of atoms results in degenerate angular momentum eigenstates.
As a result, the dipole-dipole interaction can lead to strong mixing of all degenerate or nearly-degenerate many body states.

In this paper we investigate the effect of such degeneracies on the eigenstate delocalization properties of an assembly  of $N$ particles. 
As a paradigmatic system we take an assembly of Rydberg atoms. 
These interact over micron-scale distances due to their large dipole moments, and furthermore these interactions can be tuned across several orders of magnitude by changing the principle quantum number. 
 Nearly arbitrary arrangements of atoms with relative distances on the order of a few micrometers are possible. 
The high degree of controllability makes the Rydberg assemblies perfect systems to investigate delocalized excitonic states \cite{BrBaLa16_152001_,AnVeGa98_249_,MoeWueAt11_184011_,PaGoTr16_164001_,GueScRo13_954_,AbEiEi20_124003_,AbEiEi20_193401_,ScWeBu14_63415_}.
We recently studied the eigenstates of a Rydberg gas without considering degenerate sublevels, and found that many eigenstates are delocalized over a considerable fraction of the Rydberg atoms  \cite{AbEiEi20_193401_,AbEiEi20_124003_}. 

\begin{figure}
\figd{\includegraphics[width=0.95\columnwidth]{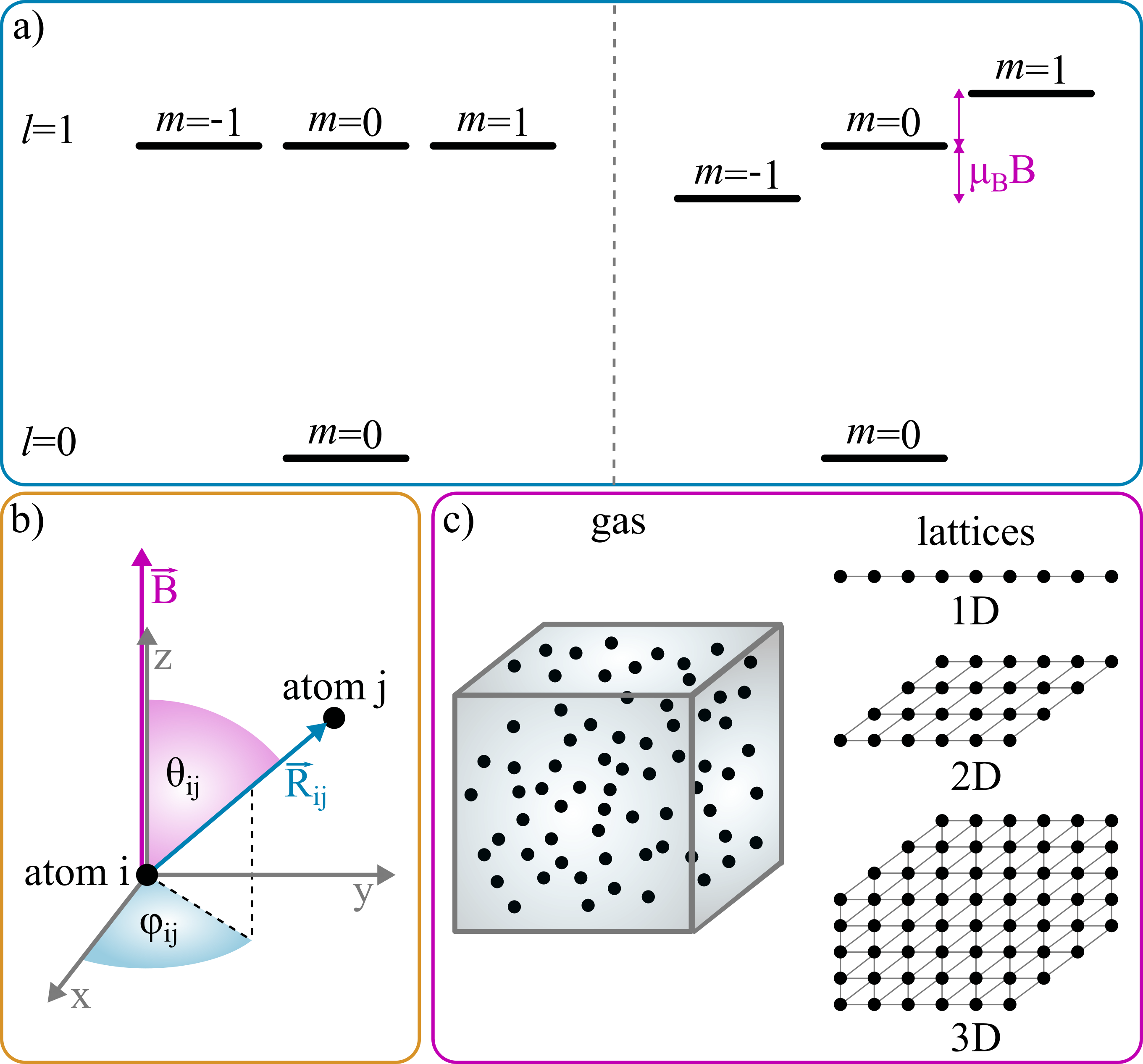}}
\caption{\label{fig:intro} (a)  Energy structure of a single Rydberg atom in the relevant subspace without (left) and with (right) an applied magnetic field. The $l = 0$, $m=0$ levels are set to be at zero energy while the $l=1$, $m=0$ state sits at energy $\epsilon$. 
(b) Definition of the angles that enter the interaction matrix elements. 
The angles $\theta_{ij}$ is defined with respect to the quantization axis, while the angle $\phi_{ij}$ is given with respect to an arbitrarily chosen $x$-axis. 
When a magnetic field is present, we choose the quantization axis $\vec{z}||\vec{B}$. 
(c) We structure our study around four generic arrangements: a three-dimensional random gas and one-, two-, and three-dimensional regular lattice arrangements. 
}
\end{figure} 

The influence of degenerate sublevels can be modified by the introduction of a static magnetic field.
In the strong field limit,  the magnetic sublevels are separated by Zeeman splittings exceeding the interaction strengths, resulting in the non-degenerate system studied previously; by varying the magnetic field we can therefore study the effect of these degeneracies in a controlled fashion. 
We find that the delocalization is enhanced by the degeneracies at zero field, resulting in even larger delocalization lengths. 
Surprisingly, these grow even further at small but non-zero magnetic fields, before reducing to the $B\to\infty$ limit studied in Refs.~\cite{AbEiEi20_193401_,AbEiEi20_124003_} at higher fields. 
To  better understand the origin of these observations we use one-, two-, and three-dimensional lattice arrangements of the  atoms \cite{browaeys2020many,barredo2018synthetic,hollerith2019quantum,de2019defect,madjarov2020high}
 to systematically probe delocalization in systems ranging from the case we previously studied -- corresponding to a disordered three-dimensional lattice with fractional filling -- to fully structured or low-dimensional systems.  
By applying also a magnetic field as we transition from regular to irregular atomic positions we can obtain further insight into the delocalization properties of this system. 

\section{Interacting Rydberg atoms}
\label{sec:2}
The role of atomic degeneracies in delocalization can be clearly studied using the spin-independent Rydberg states $\ket{\nu,l,m}$, where $\nu$ denotes the principal quantum number, $l$ the orbital angular momentum, and $m$ the corresponding magnetic quantum number.
The simplest case involves interacting s- and p-states, i.e.\ $l=0$ and $l=1$, with the same $\nu$. 
Without loss of generality we choose the $m$ quantization axis to be the same for all atoms. 
There are two manifolds of states for each atom:
\begin{align}
 \ket{\uparrow,m}& \leftrightarrow \ket{p,m} \quad \mathrm{with}\ m=0,\ \pm1 
 \\
 \ket{\downarrow,m}& \leftrightarrow \ket{s,m}\quad \mathrm{with}\ m=0.
\end{align}
We set the energy of the s-state, which does not depend on magnetic field strength,  to be the reference (zero) energy. 
The p-state energies depend linearly on an applied magnetic field via the Zeeman shift,
\begin{equation}
\label{eq:Zeeman}
\epsilon_{m}(B)=\epsilon+ \mu_{\rm B}m B,
\end{equation}
where $\epsilon$ is the energy difference between the field-free p- and s-states.
The level structure of our effectively two level system is shown in Fig.~\ref{fig:intro}(a) and (b). 

We consider an interacting system of $N$ of these two-level atoms, described by the Hamiltonian
\begin{equation}
\label{eq:ham}
{H}=\sum_{\alpha=1}^N  H^{(\alpha)} +\sum_{\alpha=1}^N\sum_{\beta<\alpha}  V^{(\alpha,\beta)},
\end{equation}
where $H^{(\alpha)}$ denotes the Hamiltonian of particle $\alpha$ and $V^{(\alpha,\beta)}$ is the dipole-dipole interaction between the atoms,
\begin{equation}
\label{eq:DipolDipole}
{V}^{(\alpha,\beta)}
= \frac{{\vec\mu}_{\alpha}\cdot {\vec\mu}_{\beta}}{R_{\alpha,\beta}^3} - 3 \frac{({\vec\mu}_{\alpha} \cdot \vec{R}_{\alpha,\beta})({\vec\mu}_{\beta}\cdot \vec{R}_{\alpha,\beta})}{R_{\alpha,\beta}^5}.
\end{equation}
 Here, $\vec{R}_{\alpha,\beta}$ is the distance vector between the two particles and ${R}_{\alpha,\beta}$ denotes its magnitude.

We are interested in the situation when there is one excitation in the system. Consequently, we choose basis states with one atom excited to the p-state and the remaining atoms still in the s-state. 
We denote these states
\begin{equation}
\label{eq:basis}
\ket{j, m_j}\equiv \ket{s,0}\cdots \ket{p,m_j}\cdots \ket{s,0},
\end{equation}
where $j$ identifies the atom which is excited to the p-state. 
The matrix elements of the Hamiltonian  (\ref{eq:ham}) in this basis are then given by
\begin{align}
\label{eq:full_rydberg_Ham}
\bra{j,m_j} H \ket{i,m_i}=& \delta_{ji}\delta_{m_jm_i} \epsilon_{m_j}(B)+ \frac{\mu_{sp}^2}{R_{ji}^3} M_{j,i}^{m_j,m_i}.
\end{align}
Here we have introduced the transition dipole moment $\mu_{\rm sp} =\bra{\nu, l=0}  r \ket{\nu, l=1} $ and a matrix element encoding the relative orientation of the atoms with respect to one another and with respect to the quantization axis,
\begin{align}
\label{eq:M_mat_elements}
M_{i,j}^{0,0}=& \dfrac{1-3\cos^2{\theta_{ij}}}{3}
\\
M_{i,j}^{+1,+1}=M_{i,j}^{-1,-1}=& - \frac{M_{i,j}^{0,0}}{2}\\
M_{i,j}^{-1,0}=& \dfrac{e^{-i\phi_{ij}}}{\sqrt{2}}\cos{\theta_{ij}}\sin{\theta_{ij}}\\
M_{i,j}^{+1,0}=&- M_{i,j}^{-1,0}\\
M_{i,j}^{-1,+1}=&\dfrac{e^{-2i\phi_{ij}}\sin^2{\theta_{ij}}}{2}.
\label{eq:M_mat_elements_end}
\end{align}
Interchanging indices results in complex conjugation.
 The angle between  the quantization axis and the distance vector  $\vec R_{ij}$ is $\theta_{ij}$,  and $\phi_{ij}$ is the azimuthal angle between the x-axis and the projection of the distance vector onto the x-y plane (see Fig.~\ref{fig:intro}(c)).

\begin{figure*}[th]
\figd{\includegraphics[width=17cm]{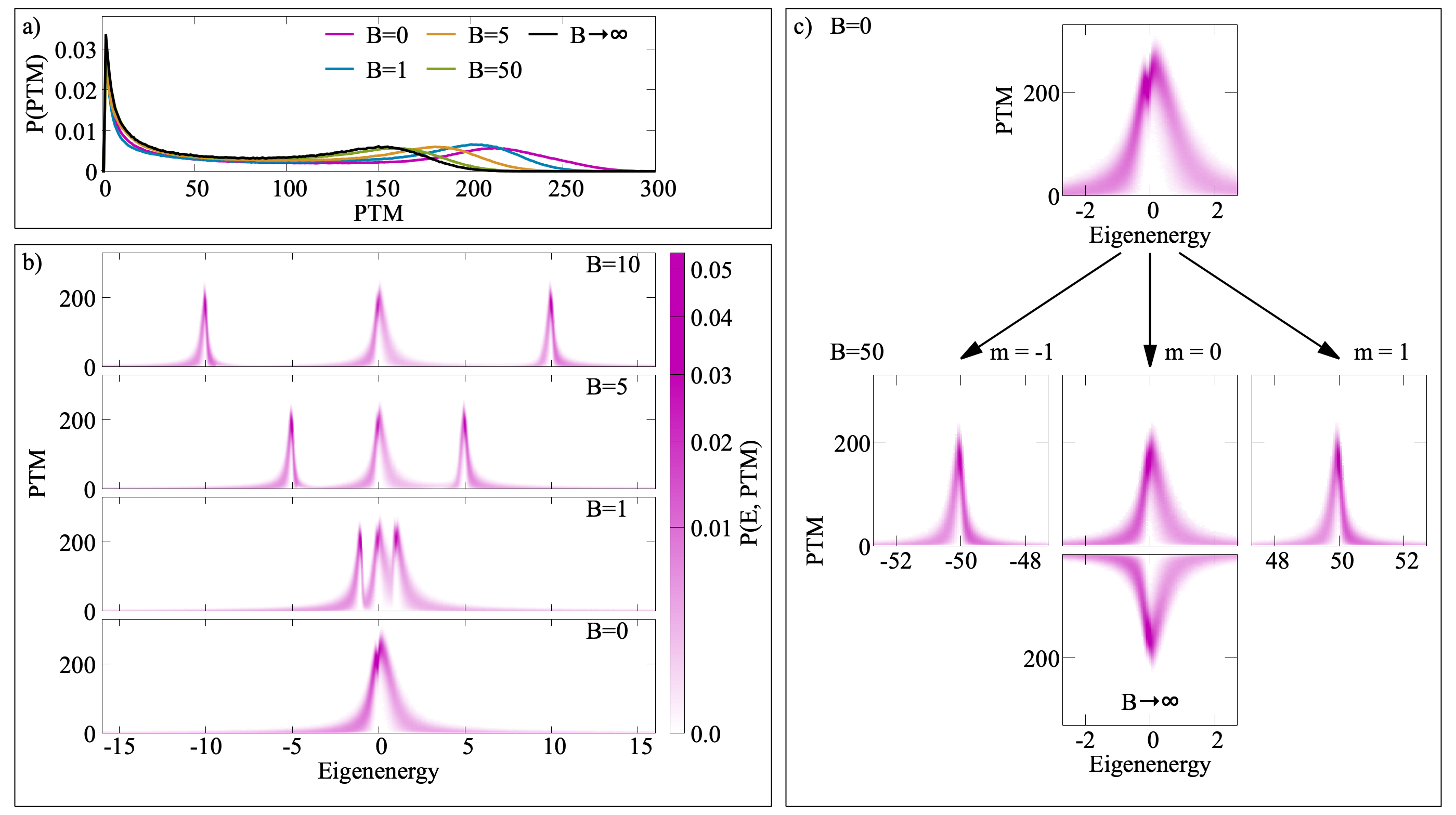}}
\caption{\label{fig:deloc_gas}
The frozen Rydberg gas. All plots are for $N= 1000$ Rydberg atoms averaged over $\sim 10^3$ realizations. (a) Probability densities for the PTM measure, displayed for several magnetic field strengths. 
The  black curve gives the $B\to\infty$ reference value, obtained by setting the coupling between $m$-levels to zero.
(b) Probability densities to find a certain PTM value for an eigenstate with a certain energy, for the same magnetic fields as in panel (a). 
(c) Comparison between the $B=0$ and  $B=50$ limiting cases. 
The energies are given in units of $E_{\rm ref}={\mu_{sp}^2}/{3a_{\rm ref}^3}$ where  $a_{\rm ref}=({3}/{4\pi N})^{1/3} L$ is the Wigner-Seitz radius. 
The zero of energy is at the energy $\epsilon$ of the non-interacting atoms, which is introduced in Eq.~(\ref{eq:Zeeman}).
}
\end{figure*}

\section{Eigenstates and delocalization measure}
\label{sec:3}
The eigenstates $\ket{\psi_\ell}$ and eigenenergies $E_\ell$ follow from the time-independent Schr\"odinger equation
\begin{equation}
\label{eq:Schroedinger}
{H}\ket{\psi_\ell}=E_\ell \ket{\psi_\ell}.
\end{equation} 
In the basis (\ref{eq:basis}) the eigenstates can be written as
\begin{equation}
\label{eq:eigenstate}
\ket{\psi_\ell}=\sum_j\sum_{\mTuple{m}} c_{j,\mTuple{m}}^{(\ell)} \ket{j,\mTuple{m}}.
\end{equation}
The absolute square of the coefficients $c_{j,\mTuple{m}}^{(\ell)}$ is the probability to find the excitation on particle $j$ in the specific state $\ket{\uparrow,m_j}$.
We obtain the eigenenergies $E_\ell$ and the eigenstate coefficients $c^{(\ell)}_{j,m_j}$ (c.f.~Eq.~(\ref{eq:eigenstate}) and Eq.~(\ref{eq:basis})) by diagonalization of a matrix with matrix elements given by  $\bra{j,m_j} H \ket{i,m_i}$  of Eq.~(\ref{eq:full_rydberg_Ham}).

We are interested in the overall delocalization of the excitation,  roughly corresponding to the number of atoms which participate in a given eigenstate. 
This is given by the probability that a particle is in the $\uparrow$-manifold. 
Since the decomposition into individual $m$ levels is irrelevant to the overall excitation delocalization,  we sum over these levels to obtain the probability that the excitation is on particle $j$, 
\begin{equation}
\label{P^(ell)}
P^{(\ell)}_j
=\sum_{\mTuple{m}}
|c_{j,\mTuple{m}}^{(\ell)} |^2.
\end{equation}
A convenient measure of delocalization can be obtained by counting the number of atoms involved in a state $\ell$ that have an excitation probability larger than a chosen threshold $P_{\rm thresh} $. 
We will refer to this as the ``population threshold measure" (PTM), 
\begin{equation}
\mathcal{N}_{\rm PTM}^{(\ell)}=\sum_j \Theta(P_j^{(\ell)}-P_{\rm thresh}),
\end{equation}
where $\Theta$ denotes the Heaviside step function.
We use $P_{\rm thresh}=1/N$; this gives a PTM limit of $N$ in a fully delocalized, equally distributed, state, and $1$ for a state localized on a given atom. 

In our previous work we used the so-called ``coherence" measure to quantify the delocalization of the excitation.
Since the PTM measure works directly with the populations, it is more suitable for the present suituation where we are not interested in the coherence properties of the reduced density matrix. 
In the Supplemental Material of Ref.~\cite{AbEiEi20_193401_} we compared these measures for the case without $m$-levels, and observed that they are essentially proportional.

\section{The frozen Rydberg gas}
\label{sec:4}
In a frozen Rydberg gas the atoms are randomly distributed within a certain volume and, due to the typical ultracold laboratory conditions and relevant time scales, remain motionless during the course of excitation and measurement of delocalized states.
For a representative study we consider $N=1000$ Rydberg atoms with random positions placed uniformly inside a cubic volume with length $L$; the results are nearly independent of boundary conditions  \cite{AbEiEi20_193401_,AbEiEi20_124003_}.
It is convenient to use the Wigner-Seitz radius $a_{\rm ref}=(3/4\pi N)^{1/3} L$  as the unit of distance, 
and for the unit of energy based on typical dipole-dipole interaction strengths we use $E_{\rm ref}={\mu_{sp}^2}/{3 a_{\rm ref}^3}$.
We average over $10^3$ independent random gas realizations. 

In Fig.~\ref{fig:deloc_gas} the dependence of the PTM measure on the magnetic field strength is shown.
Fig.~\ref{fig:deloc_gas}(a) shows the probability density for finding a certain PTM for several magnetic field strengths. 
For all magnetic field strengths there is a large fraction of states with PTM on the order of {150-200}, i.e. the delocalization is spread over nearly 20\% of the atoms in the gas. 
The PTM distribution is shifted towards larger values for all finite B-fields in comparison to the $B\to\infty$ case, to which they converge.
 The peak at low PTM stems from clusters -- dimers, trimers, etc.-- formed from strongly-interacting atoms in relatively close promixity. 
These cluster states decouple from the system, leaving a residual gas with more homogenous inter-particle interactions, which in turn lead to large extended states \cite{AbEiEi20_193401_}. 

\begin{figure}
\figd{\includegraphics[width=8cm]{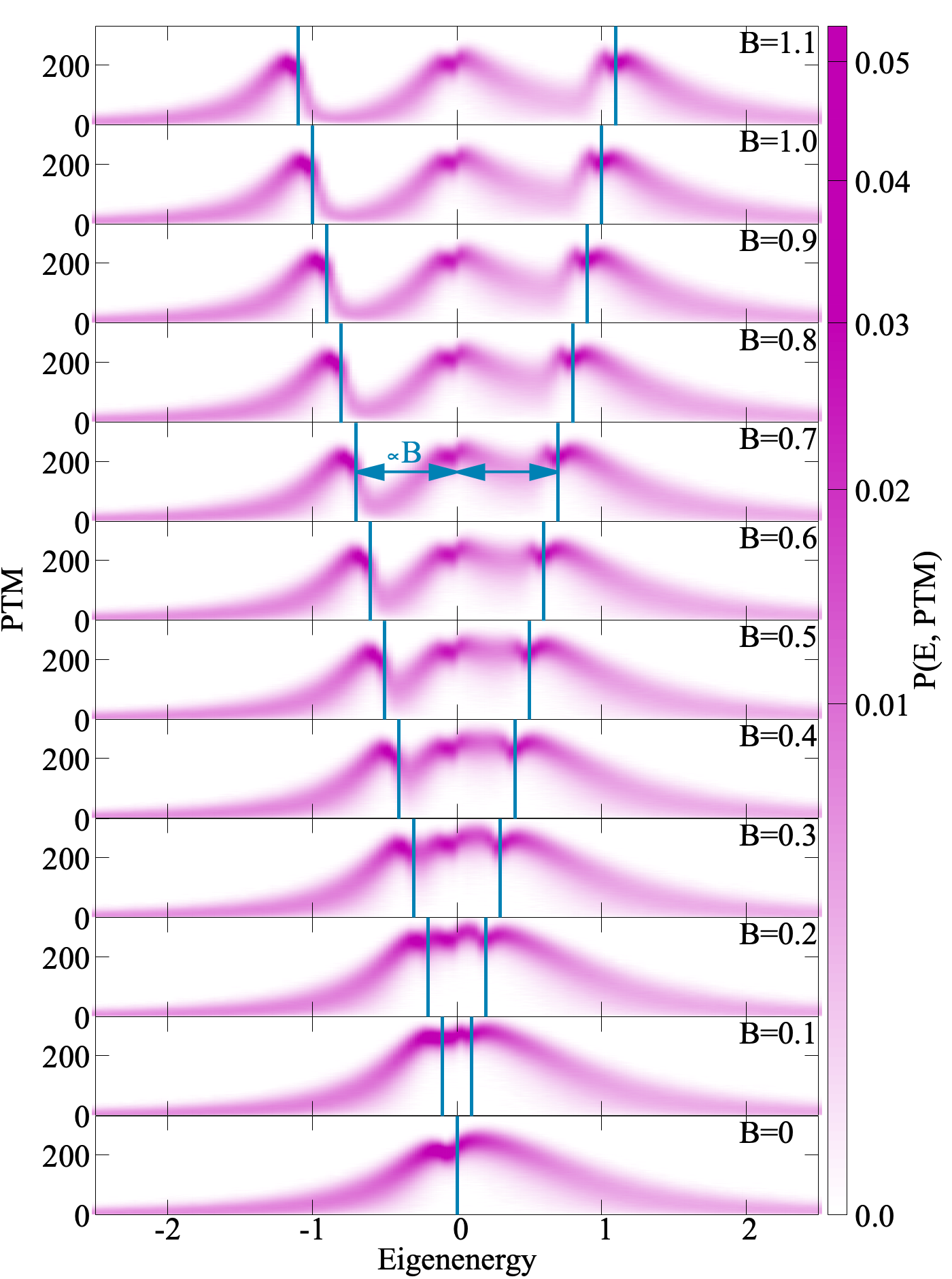}}
\caption{\label{fig:density_deloc_gas}
Same as Fig.~\ref{fig:deloc_gas}(b), but for small magnetic fields.
The vertical bars indicate the positions of the Zeeman energies, $\pm m B$.
}
\end{figure}

This interpretation is supported by the energy-resolved PTM distributions shown in Fig.~\ref{fig:deloc_gas}(b).
These reveal immediately that the delocalized states cluster around zero energy while the cluster states with low PTM are primarily found in the wings of this distribution. 
One sees additional structure in the PTM distributions: at $B=0$ the distribution has an asymmetric double-peak structure and is even broader than in the $B\to\infty$ case.  
For intermediate $B$ the distribution splits into three peaks whose centers follow the Zeeman energies proportional to the magnetic field strength $B$.
For $B \gtrsim 10$ the three well-separated peaks each have states with only a single $m$ value. 
These peaks have a similar shape, but different width and ``orientation".
While $B=10$  is not quite sufficient to reach the asymptotic $B\to\infty$ value, $B=50$,  shown in panel (c), is. 
One clearly sees that the $m=\pm 1$ peaks are mirror images of the $m=0$ peak with half the width, features which result from the form of the interaction matrix. 
In the $B\rightarrow \infty$ limit the off-diagonal couplings in $M$ can be ignored, and thus the Hamiltonian  separates into three blocks, with energies $m B$ on the diagonal and off-diagonal elements given by $M_{i,j}^{0,0}$, $M_{i,j}^{-1,-1}$ and $M_{i,j}^{+1,+1}$.   
From Eqns.~(\ref{eq:M_mat_elements})--(\ref{eq:M_mat_elements_end}) one sees that $M_{i,j}^{-1,-1}$ and $M_{i,j}^{+1,+1}$ have the same sign and magnitude, but a different sign and half the magnitude of the $M_{i,j}^{0,0}$ interaction.
It is clear that the PTM distributions for all three blocks are identical, since the interactions are proportional. 
Mirrored below the $m=0$ distribution we show the asymptotic $B\rightarrow \infty$ result \cite{AbEiEi20_193401_}. 
It is identical to the $B = 50$ $m =0$ distribution, confirming the validity of the results of Ref.\cite{AbEiEi20_193401_}, since the non-degenerate regime is reached in the limit of moderately high magnetic fields. 
Specifically, at typical Rydberg densities the interaction strength is on the order of a few MHz, which requires a magnetic field on the order of 10G to reach the separated $m$-level regime. 

\begin{figure*}[t]
\figd{\includegraphics[width=15cm]{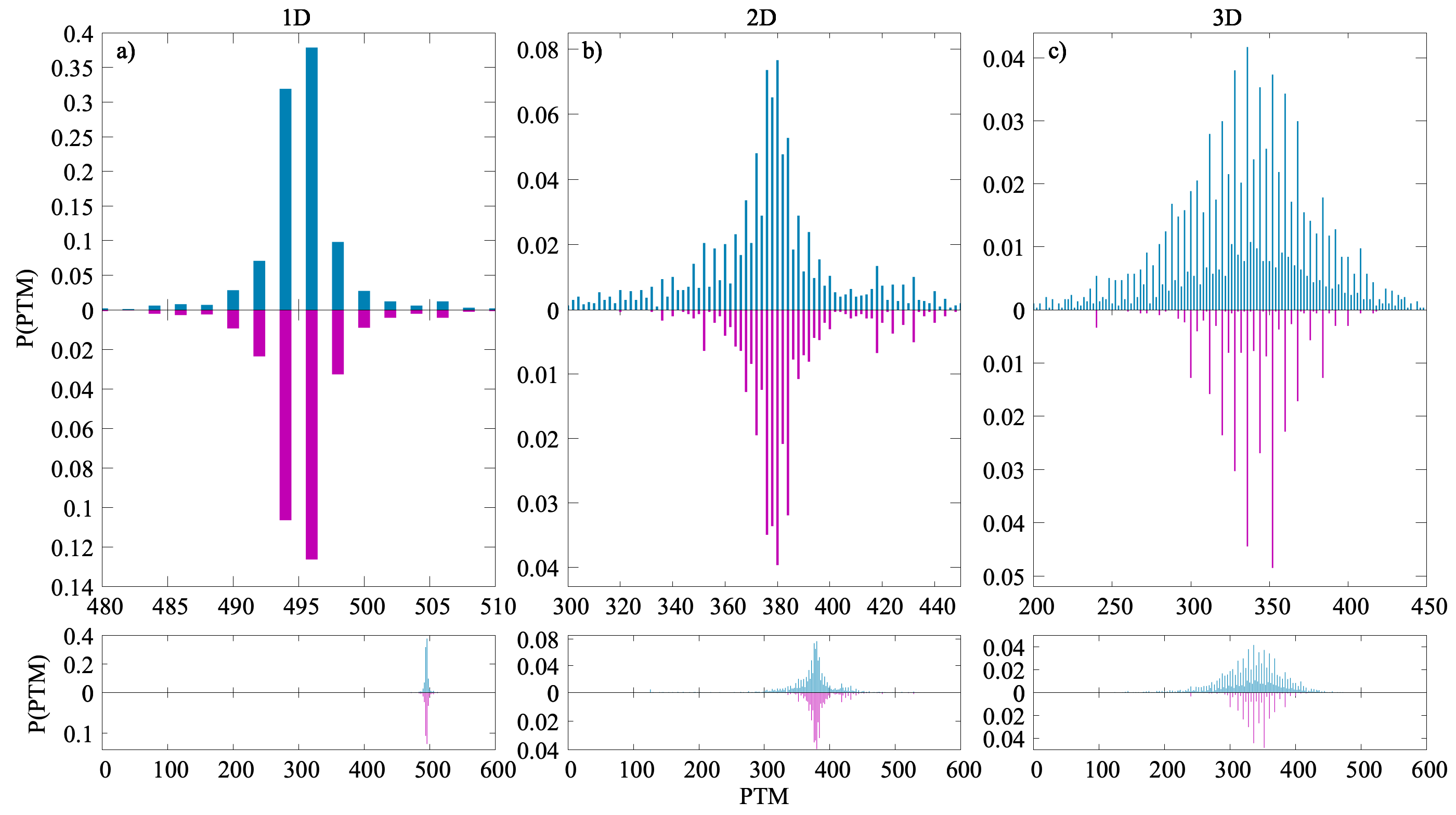}}
\caption{\label{fig:lattices_deloc}Distribution of PTM for (a) a 1D chain, (b) a 2D and (c) a 3D lattice.  In all three cases we have roughly the same number of atoms as in the frozen gas case (990 atoms in the 1D case ($N_x=990$), 992 atoms in the 2D case ($N_x=32$, $N_y=31$), and 990 atoms in the 3D case ($N_x=11$, $N_y=10$, $N_z=9$), and the quantization axis is set parallel to the magnetic field, pointing along the z-direction.
The blue bars are the case without magnetic field ($B=0$) and the magenta (mirrored) bars are the infinite B-field limit. The bottom row shows the upper plots on the same x-axis.
}
\end{figure*}

To study in more detail the splitting of the asymmetric $B=0$ distribution into three peaks, in Fig.~\ref{fig:density_deloc_gas} we focus on the region $B\leq 1.1$.
Surprisingly, the apparently monotonic decrease in delocalization extent with increasing $B$ seen in Fig.~\ref{fig:deloc_gas} does not hold all the way to $B = 0$, and in fact the largest delocalized states are seen for small but non-zero fields with partially lifted degeneracy. 
For $B\lesssim 1$ the magnetic field is too weak to separate the three distributions fully, resulting in a complicated distribution with several maxima culminating in a clearly emerging triple peak structure as $B$ grows to $\approx 1$. 
 The peak centered at zero detuning has a {similarly} asymmetric form as it does when $B=0$,  but with a smaller width and reduced maximum value. 
The right peak ($m=+1$) has a strongly asymmetric shape and exhibits a similar double peak structure as the $m=0$ peak.
In contrast, the left peak ($m=-1$) has no double peak structure and its asymmetry is mirrored with respect to the $m=+1$ peak. 
Recall that in the large B-field case these two peaks are identical. 

From these calculations we see that the degenerate sublevels increase the extent of delocalization in and add additional structure to the random Rydberg gas. 
Their interplay with a magnetic field leads to complicated behavior, even including an increase in the delocalization length at small magnetic fields and hence a small lifting of the degeneracy, which eventually converges to the non-degenerate $B\to\infty$ case studied in Refs. \cite{AbEiEi20_193401_,AbEiEi20_124003_}.
The ability to use an applied magnetic field to subtly tune both the level structure and delocalization properties of such a gas provides one way of comparing the degenerate and non-degenerate limits. 
{To better understand the effect of the degenerate atomic transitions we now consider {different} lattice arrangements.}

\section{Atoms arranged on a lattice}
\label{sec:5}
\subsubsection{One-dimensional chain}
We consider $N$ equidistant atoms placed in a one-dimensional (1D) lattice. 
We first note that the angle between the quantization axis and the vector $\vec{R}_{ij}$ is the same for all pairs of atoms, i.e.~$\theta_{ij}\equiv\theta$ and $\phi_{ij}\equiv \phi$ for all $i$ and $j$.
The Hamiltonian can therefore be simplified to 
\begin{equation}
\label{eq:Ham,chain1}
H= \sum_{i,j} \sum_{m,m'}\Big[\epsilon_m(B)\delta_{m,m'}\delta_{ij}+ \frac{\mu_{\rm sp}^2}{R_{ij}^3}  M_{m,m'} \Big]\ket{i,m}\bra{j,m'},
\end{equation}
where $M^{m,m'}=M_{i,j}^{m,m'}$  are independent of atom indices. 
This has far-reaching consequences. 
{As shown in appendix \ref{app:1D_analytic},} the PTM distribution is actually independent of the direction and strength of the magnetic field,  in pronounced contrast to the three-dimensional gas, and even though the eigenvalues depend on the magnetic field. 
This holds even when the atoms are not placed equidistantly. 
The PTM for $N=990$ is shown in Figure \ref{fig:lattices_deloc}(a) for $B=0$ and $B\rightarrow \infty$ (mirrored below).
The PTM distribution is centered around $N/2$, in excellent agreement with the analytic estimates discussed in appendix \ref{app:estDeloc}.

\subsubsection{Two-dimensional lattice}
We now place the atoms in a two-dimensional (2D) rectangular lattice in a plane perpendicular to the quantization axis so that $\theta_{ij}=\pi/2$ for all $i,j$.
In this case  $M_{i,j}^{\pm 1,0}=M_{i,j}^{0,\pm 1}=0$,  and thus the $m=0$ subspace decouples from the $m=\pm 1$ states.
Within the $m=0$ subspace the interaction {${\mu_{\rm sp}^2}/3{R_{ij}^3}$} is isotropic,  and was previously studied without the lattice arrangement in Ref.~\cite{AbEiEi20_124003_}. 
The $m=0$ PTM distribution is independent of $B$.
In Figure \ref{fig:lattices_deloc}(b) we contrast the full PTM distribution for the case $B=0$ ( top panel) with the $B\rightarrow \infty$ case (bottom panel), with the magnetic field perpendicular to the lattice. 
The two distributions are not equal due to the $m=\pm 1$ states present only in the $B = 0$ case.
 By comparing the differences between the two mirrored distributions, it is apparent that these states both increase the number of highly delocalized states and give rise to several somewhat more localized states which are completely absent in the $B\to\infty$ case. 
Unlike in the 1D case,  in the 2D case $B\ne 0$ both the eigenstates and the eigenenergies depend on the magnetic field orientation. 

\subsubsection{Three-dimensional lattice}

A three dimensional lattice (3D) bears the closest resemblance to the frozen gas. 
Fig. \ref{fig:lattices_deloc}{(c)} shows that the coupling between $m$ levels still has only a a small impact on the delocalization, which, as in the 1D and 2D cases is characterized by PTM values around $N/2$, although in 3D the distribution is broader. 
In marked contrast to the 1D and 2D cases there are now states with larger PTM values than in the infinite B-field limit. 
We therefore see that the main finding of the previous section, that the degenerate sublevels at $B = 0$ lead to larger delocalization than in the non-degenerate $B\to\infty$ limit, only occurs for the 3D arrangement. 

There are several differences between the 3D-lattice and the random gas PTM distributions. 
Most notably, the lattice PTM distribution consists only of a single-peaked and relatively narrow distribution centered at a high PTM value of approximately $N/3$; in contrast, the random gas case exhibits a very broad distribution with two major peaks at PTM values around 2 and at $\approx N/5$, 
First,  there is a broad distribution of {PTM} values, with peaks at 0 and $\sim 200\approx N/5$. 
These differences are explained by the absence of strongly interacting clusters in the lattice. 

\begin{figure}
\figd{\includegraphics[width=8cm]{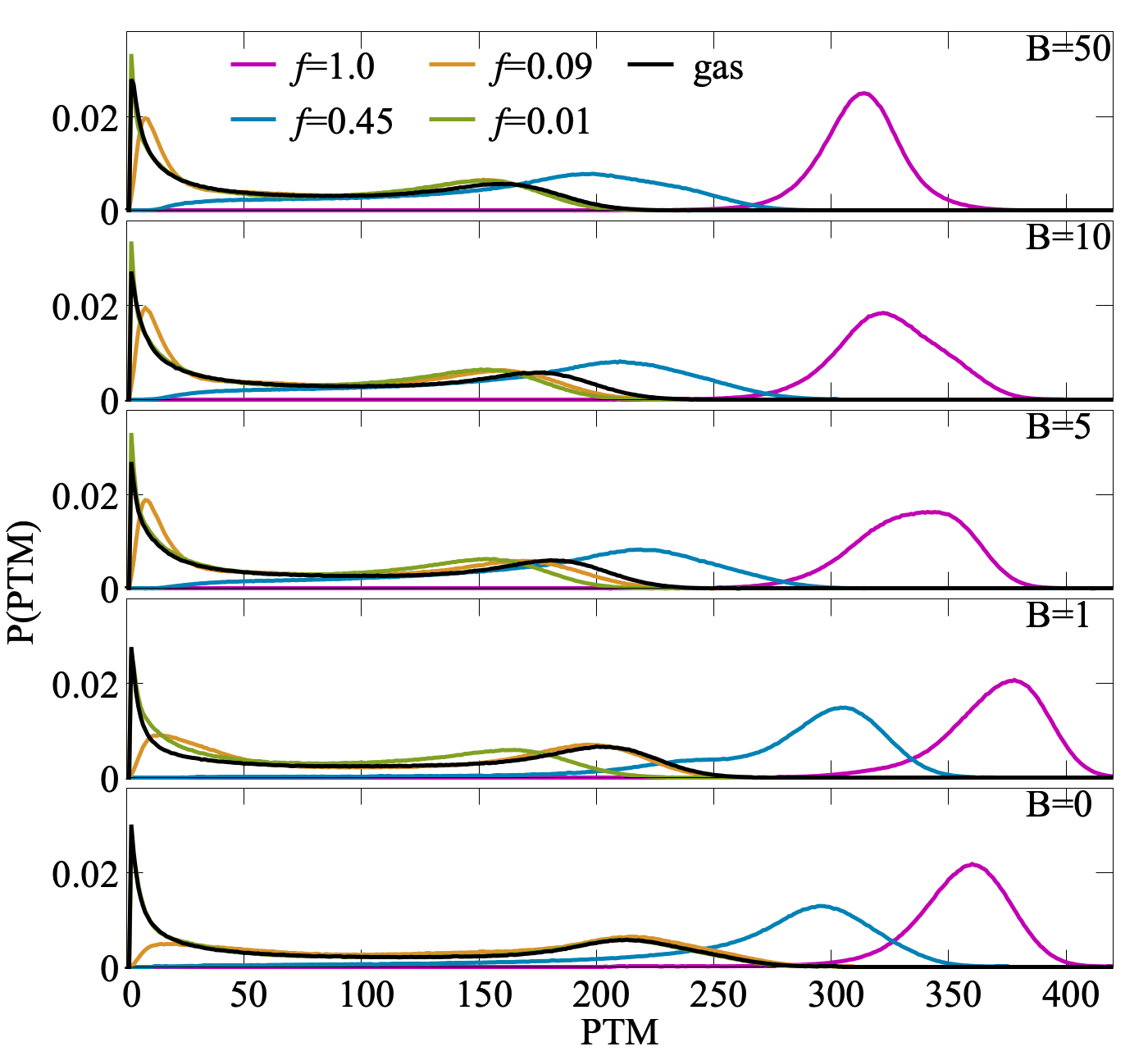}}
\caption{\label{fig:LatticeToGas} Transition from lattice to gas. Comparison of a 3D lattice with different filling fractions {(numbers and colors provided in the upper panel)} to the frozen Rydberg gas (black) for different strengths of the magnetic field (provided in the panels). For all cases we fix the number of atoms at $N=990$. For the lattice we apply $5\%$ uniform fluctuations around the perfect lattice positions.
}
\end{figure}

To study the transition from the 3D lattice case to the frozen gas, we introduce now a lattice with a partial filling fraction $f$, which introduces clustering effects into the lattice. 
For a given $f$ we adjust the size of the lattice such that we always have the same total number of atoms in the system. 
In Fig.~\ref{fig:LatticeToGas} we show the  PTM distribution for several filling fractions and magnetic fields.
To smoothen the distribution with high filling fraction we added a small ($5\%$ of the lattice constant) disorder in the position of the atoms around the lattice positions.
As expected, the peak in the PTM distribution starts to broaden and shifts to smaller values as $f$ decreases.
For $f = 0.09$ a peak at small ($<50$) PTM values develops, becoming more pronounced at higher $B$ values. 
We note that for $B=0$ and for large magnetic field ($B=50$) the distribution at small filling fraction ($f=0.01$) agrees nearly perfectly with the frozen gas distribution (shown as black line).  
Curiously, at intermediate magnetic field strengths the distributions with small filling fractions have peaks at smaller PTM values than the frozen gas case, suggesting here also a complicated interplay between the degenerate sublevels and their anisotropic interactions, the lattice structure, relative alignment of the lattice and magnetic field, and magnetic field strength. 

\section{Conclusions}
In this paper we have explored the influence of degenerate sublevels on the extent of single exciton state delocalization in a Rydberg atom assembly. 
Taking the degenerate sublevels into account, we find larger delocalization than for the separate $m$-level manifolds in the three-dimensional system. 
The extent of this delocalization can be controlled via an external magnetic field, which tunes the system between the degenerate $m$-level and decoupled $m$-level limits in the $B = 0$ and $B\to\infty$ limits, respectively. 
In one and two dimensions the inclusion of $m$-level degeneracies does not lead to much larger delocalization lengths, and we find that the magnetic field orientation does not influence the delocalization properties of the system at all in one dimension, and only weakly in three dimensions. 
The situation in two dimensions is very different; the strong dependence on the direction of the external magnetic field here is important since this system mimics the situation of regular 2D molecular arrangements on surfaces. For example, for PTCDA-molecules on a KCl surface, all molecular transition dipoles are oriented along the diagonal of the lattice \cite{EiMaPa17_097402_,MPaEi13_044302_}, an interesting situation which can be simulated in the Rydberg assembly. 

Our choice of $s$ and $p$ states to study exciton delocalization was made in order to introduce a tractable level of degeneracy to the system. 
If, instead, we had chosen $p$ and $d$ states as $\downarrow$ and $\uparrow$ states, respectively,  then in addition to a small increase in the degeneracy of $\uparrow$ states, we would introduce doubly degenerate $\downarrow$ states, leading to an exponential growth in the number of aggregate basis-states.
Such a case is challenging to treat numerically, but is the typical situation in molecular systems.
Inclusion of Rydberg fine structure also leads to such a scenario for all possible states, since the spin degree of freedom leads  even to a doubly degenerate $s$ state. 
Exploration of this physics would be both theoretically interesting along the lines of choosing other degenerate states, but is also necessary in order to treat realistic experimental conditions \cite{lippe2020experimental,lienhard2020realization}

\begin{acknowledgments}
We acknowledge funding from the DFG: grant EI 872/4-1
through the Priority Programme SPP 1929 (GiRyd).
AE acknowledges support from the DFG via a Heisenberg fellowship (Grant No EI 872/5-1). MTE acknowledges support during early stages of the research from the Alexander von Humboldt Stiftung.  CWW acknowledges support from the Max-Planck Gesellschaft via the MPI-PKS Next Step fellowship.
\end{acknowledgments}

%\bibliographystyle{apsrev4-1}
%\bibliography{Gas_Ryd_Alex}

\appendix
\section{\label{app:1D_analytic}Eigenstate structure in the 1D chain}
The Hamiltonian of a 1D-chain, Eq.~(\ref{eq:Ham,chain1}), can be written in matrix form as
\begin{equation}
\label{eq:Ham,chain2}
H(B)= \underline{\epsilon(B)} \otimes \underline{\underline{I}} + \underline{M} \otimes \underline{\underline{V}}
\end{equation}
Here a single underbar denotes a $3\times 3$ matrix and a double underbar denotes a $N \times N$ matrix.
The symbol $\otimes$ denotes the Konecker product between matrices.
These matrices are given by
\begin{equation}
\underline{\epsilon(B)}
=\left(\begin{array}{ccc}
- \mu_{\rm B} B&0 &0 \\
0 &0&0\\
0& 0& +\mu_{\rm B} B
\end{array}
\right)
\end{equation}
\begin{equation}
\underline{M}=
\small
\left( {\begin{array}{ccc}
		\dfrac{3\cos^2{\theta}-1}{6}&
		 \dfrac{e^{-i\phi}}{\sqrt{2}}\cos{\theta}\sin{\theta}&
		\dfrac{e^{-2i\phi}\sin^2{\theta}}{2}\\

		\dfrac{e^{i\phi}}{\sqrt{2}}\cos{\theta}\sin{\theta}&
		 \dfrac{1-3\cos^2{\theta}}{3}&
		 -\dfrac{e^{-i\phi_{ij}}}{\sqrt{2}}\cos{\theta}\sin{\theta}\\

		\dfrac{e^{2i\phi}\sin^2{\theta}}{2}&
		 -\dfrac{e^{i\phi}}{\sqrt{2}}\cos{\theta}\sin{\theta} &
		\dfrac{3\cos^2{\theta}-1}{6}\\
\end{array} } \right)
\end{equation}
The matrix $\underline{\underline{V}}$ contains the elements
\begin{equation}
V_{ij}=\mu_{\rm sp}^2/R_{ij}^3
\end{equation}
and $\underline{\underline{I}}$ is the $N\times N$ unit matrix.

In a first step we can diagonalize the matrix $\underline{\underline{V}}$:
\begin{equation}
\underline{\underline{V}}\vec{a}^{(\alpha)}=E^{(\alpha)}\vec{a}^{(\alpha)}
\end{equation}
where $\alpha$ labels the $N$ eigenvectors.
Then Eq.~(\ref{eq:Ham,chain2}) can be written as
\begin{eqnarray}
H(B) [ \underline{I}\otimes \vec{a}^{(\alpha)}]&=& \underline{\epsilon(B)} \otimes \vec{a}^{(\alpha)} + E^{(\alpha)}\underline{M} \otimes \vec{a}^{(\alpha)}
\\
&=&\big[ \underline{\epsilon(B)} + E^{(\alpha)}\underline{M}\big] \otimes \vec{a}^{(\alpha)}
\end{eqnarray}
In the next step we diagonalize the $N$ $3\times3$ matrices $\underline{\epsilon(B)} + E^{(\alpha)}\underline{M}$:
\begin{equation}
\big[\underline{\epsilon(B)} + E^{(\alpha)}\underline{M}\big] \vec{b}^{\alpha,\beta}(B)= E^{\alpha,\beta} (B)\, \vec{b}^{\alpha,\beta}(B)
\end{equation}
Here $\beta$ labels the $3$ eigenvectors of each $\beta$-block.
With this we finally can write:
\begin{equation}
H(B) [ \vec{b}^{(\alpha,\beta)}(B)\otimes \vec{a}^{(\alpha)}]= E^{(\alpha,\beta)} (B)[\vec{b}^{(\alpha,\beta)}(B)\otimes \vec{a}^{(\alpha)}]
\end{equation}
We can combine the two labels $\alpha$ und $\beta$ into a single label $\ell$ and define as eigenfunctions.
\begin{equation}
\vec{c}^{(\ell)}=\vec{b}^{(\alpha,\beta)}\otimes \vec{a}^{(\alpha)}
\end{equation}
From this we can make the identification 
\begin{equation}
c^{(\ell)}_{j,m}= b^{(\alpha,\beta)}_m a^{(\alpha)}_j
\end{equation}
Since we are interested in the populations on each site (see Eq.~(\ref{P^(ell)}) we find 
\begin{equation}
P^{(\alpha,\beta)}_j
=\sum_{{m}}
|c_{j,{m}}^{(\ell)} |^2=\big(\sum_{m} |b_m^{(\alpha,\beta)}|^2 \big)\,| a_j^{(\alpha)}|^2 
\end{equation}
Since $\big(\sum_{m} |b_m^{(\alpha,\beta)}|^2 \big)=1$ we can finally write
\begin{equation}
P^{(\ell)}_j
=| a_j^{(\alpha)}|^2 
\end{equation}
From this one sees that the populations are independent of the magnetic field strength and direction.
They are given by the $m$-level independent Hamiltonian $\underline{\underline{V}}$, which corresponds to isotropic interaction.	

{{\it Alternative considerations using $B=0$:}}
With $B=0$ it is convenient to choose the quantization axis such that all coupling elements ($M_{i,j}^{0,\pm 1}$, $M_{i,j}^{\pm 1,0}$ and $M_{i,j}^{\pm 1,\mp 1}$) vanish. This happens when the quantization axis is parallel to the chain ($\theta=0$).
That means that the $m=-1$, $m=0$ and $m=+1$ states are uncoupled and the Hamiltonian has a block-diagonal form, where each block belongs to a specific $m$-state.
Each block can be diagonalized independently.
From the definition of the $M_{i,j}^{m_i,m_j}$ one sees that the three sub-blocks have the form
\begin{equation}
\label{eq:1D-chain}
H^{(m)}= \sum_{i,j} \big[\epsilon\delta_{ij}+ \frac{\mu_{\rm sp}^2}{R^3} \frac{1}{|i-j|^3} M^{(m)} \big]\ket{i,m}\bra{j,m}
\end{equation}
with $M^{(0)}=1/3$ and $M^{(\pm 1)}= - 1/6$, and where $R$ is the lattice spacing. 
Since $M^{(m)}$ is independent of the atomic position $i$ and $j$, it simply scales the interaction strength.
Therefore, one has identical eigenstates for each sub-Hamiltonian.

\section{\label{app:estDeloc} PTM estimate for the 1D chain}
To analytically estimate the extent of the delocalization we take only the nearest neighbor interaction into account.
The squares of eigenfunction coefficient which are used in the calculation of the PTM then are given by ${|c_j^{(\ell)}|^2}=2/(N+1) \sin^2(\pi j \ell /(N+1))$.
One sees that roughly half of them are larger than the threshold $P_{\rm thresh}=1/N$. 
Therefore, we expect the eigenstates to have a PTM value of  approximately $N/2$.

\end{document}